\documentclass{ws-procs9x6}
\def\Journal#1#2#3#4{{#1}{\bf #2}, #3 (#4)}
\def\HIP{{\it Heavy Ion Physics}\ }

\def\NPA{{\it Nucl. Phys.}~{\bf A}\,}
\def\NPB{{\it Nucl. Phys.}~{\bf B}\,}
\def\PLB{{\it Phys. Lett.}~{\bf B}\,}

\def\PRL{\it Phys. Rev. Lett.\ }
\def\PR{\it Phys. Rev.\ }
\def\PRD{{\it Phys. Rev.}~{\bf D}\,}
\def\PRC{{\it Phys. Rev.}~{\bf C}\,}
\def\ZPC{{\it Z. Phys.}~{\bf C}\,}
\def\BAPS{{\it Bull. Am. Phys. Soc.}\ }
\begin{document}

\title{Lepton and Photon Physics at RHIC\footnote{\uppercase{T}his work is supported by \uppercase{U.S. D}epartment of \uppercase{E}nergy under contract \uppercase{DE-AC}02-98\uppercase{CH}10886.}}

\author{M. J. TANNENBAUM}

\address{Brookhaven National Laboratory \\
Physics Department, 510c \\ 
Upton, NY 11973-5000, USA\\ 
E-mail: mjt@bnl.gov}


\maketitle

\abstracts{Results on physics at RHIC using outgoing leptons and photons will be presented from Au+Au collisions at nucleon-nucleon c.m. energies $\sqrt{s_{NN}}=130$ GeV and 
200 GeV, and from p-p collisions at $\sqrt{s_{NN}}=200$ GeV. 
Introduction and motivation will be presented both from the theoretical and 
experimental perspectives. Topics include open charm production via single 
$e^{\pm}$, $J/\Psi\rightarrow e^+ + e^-$, $\mu^+ + \mu^-$ and inclusive photon production.}

\section{Motivation}
\subsection{QCD and Dileptons}\label{subsec:QCD}

	Lepton pair production in hadron collisions has played a vital role in the development of the Standard Model of elementary particle physics in both the Electro Weak and the strong interaction sector. In Quantum Chromo Dynamics, the theory of the strongly interacting sector, the force between a quark and an anti-quark, coupled to color, is coulomb-like at short distances, leading to bound states, but is string-like at large distances, leading to confinement.\cite{cornell} 
	
	There is also the complication of 3 families of quark and lepton doublets. The heavy  quarks $c$, $b$, which couple weakly to the light quarks $u$, $d$, $s$, were discovered in the 1970's via the large branching ratios $\sim 10^{-2}$ of the lowest bound states $J/\Psi$ ($c\bar{c}$)\cite{Ting} and $\Upsilon$ ($b\bar{b}$)\cite{Lederman} to dileptons, due to the empirical `family conservation' law, compared to the $\sim 10^{-4}$ dilepton branching ratio of the light quark vector mesons $\rho^0$ and $\omega$. 
	
\begin{figure}[ht]
\begin{minipage}[b]{2.2in}
\centerline{\psfig{file=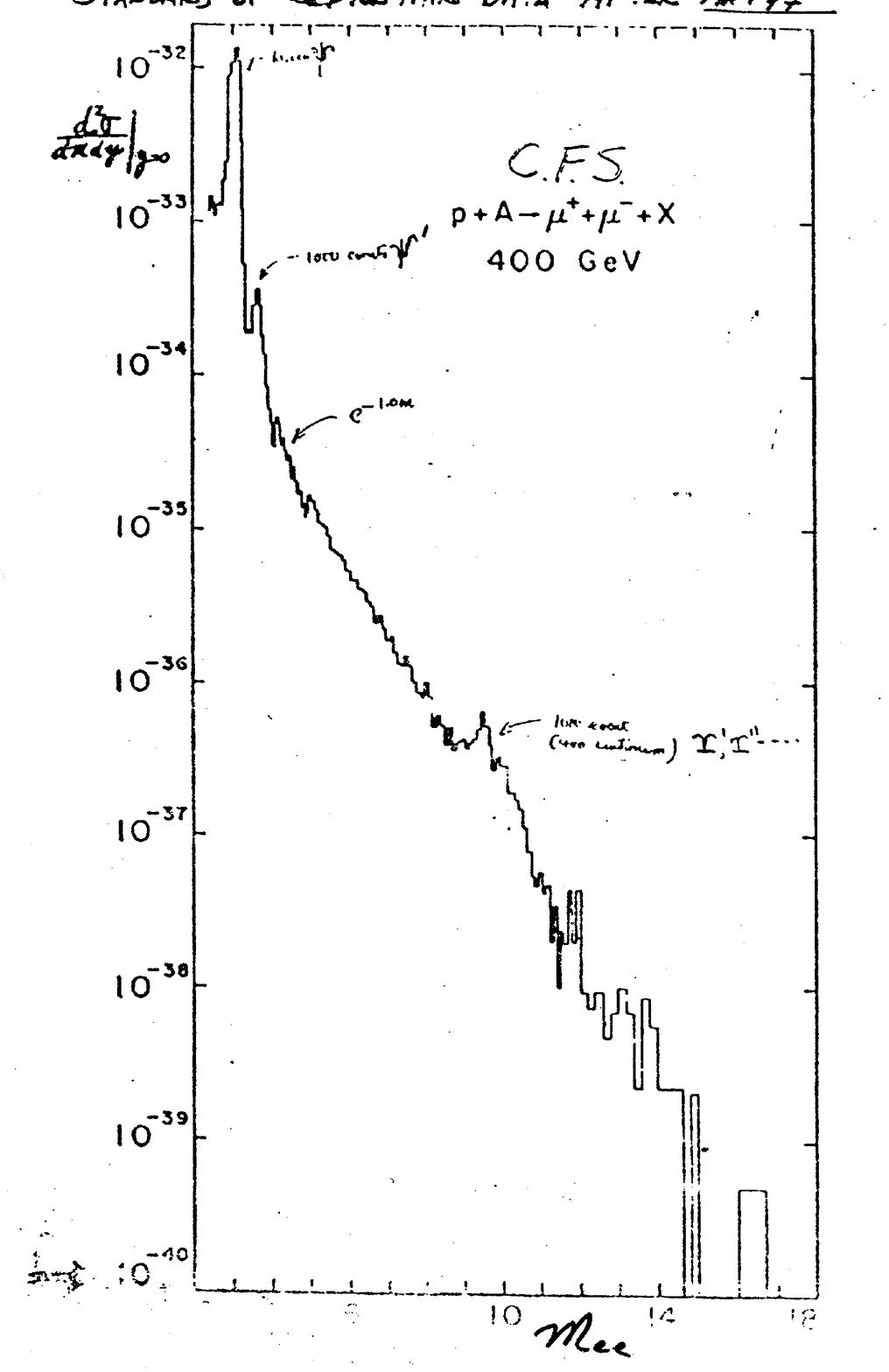,width=2.2in}}
\end{minipage}
\begin{minipage}[b]{2.2in}
\centerline{\psfig{file=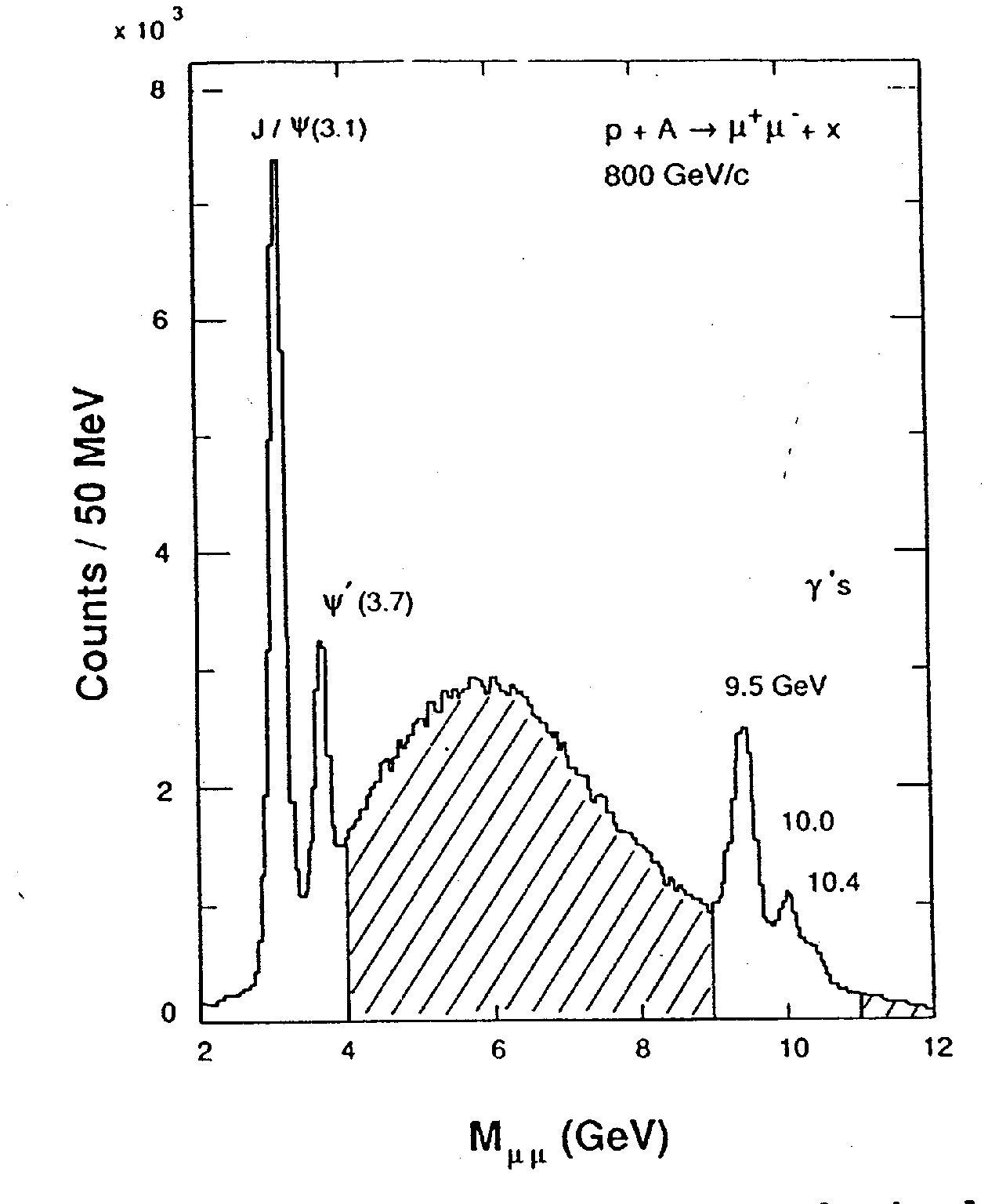,height=3in}} 
\end{minipage}
\caption[]{World-class dimuon spectra, ({left}) c. 1977\cite{Lederman},  ({right}) c. 1991\cite{E772}. }
\label{fig:cfs}
\end{figure}

	All this is beautifully illustrated in Fig.~\ref{fig:cfs} which on the left shows the cross section $d^{2}\sigma /dm dy|_{y=0}$ at mid-rapidity as a function of dimuon invariant mass,\cite{Lederman} with the bound states $\Psi$ and $\Psi^{'}$, and the $\Upsilon$ family, clearly visible upon a continuum which appears to fall exponentially as $e^{-1.0 m}$ at $\sqrt{s}=27.4$ GeV. The continuum, commonly known as Drell-Yan\cite{DY} (although discovered at the BNL-AGS by Leon Lederman and collaborators\cite{LMLDY}) is due to the constituent reaction $q+\bar{q}\rightarrow \mu^+ +\mu^-$. Fig.~\ref{fig:cfs}(right) shows a more recent raw dimuon mass spectrum\cite{E772} from p+A collisions at Fermilab illustrating the state of the art in these measurements.
	
\subsection{$J/\Psi$, $\Upsilon$ and the Quark Gluon Plasma (QGP)}\label{subsec:QGP}

\begin{figure}[ht]
\begin{minipage}[b]{2.2in}
\centerline{Success in HEP}
\centerline{\psfig{file=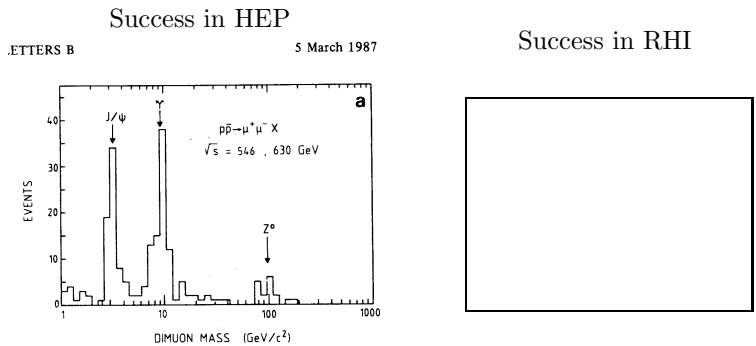,width=2.0in}}
\end{minipage}
\begin{minipage}[b]{2.2in}
\centerline{Success in RHI\hspace*{0.25in}}
\vspace*{0.25in}
\hspace*{0.25in}\begin{picture}(108,80)
\put(0,0){\framebox(108,80){}}
\end{picture}
\vspace*{0.25in}
\end{minipage}
\caption[]{``The road to success": In High Energy Physics (left) a UA1 measurement\cite{UA1} of pairs of muons each with $p_T\geq 3$ GeV/c shows two Nobel prize winning dimuon peaks and one which won the Wolf prize. Success for measuring these peaks in RHI physics is shown schematically on the right. }
\label{fig:success}
\end{figure}

	Dilepton production also figures prominently as one of the `Gold Plated' signatures for deconfinement in a Quark Gluon Plasma (QGP)---$J/\Psi$ suppression.\cite{Matsui} The attractive short range QCD potential is `Debye screened' in the QGP causing any $c\bar{c}$ or $b\bar{b}$ bound states produced in the early partonic phase of a Relativistic Heavy Ion Collision to dissolve into unbound $c$ or $b$ quarks, which later form charm or bottom mesons at the freezeout stage. The suppression is very sensitive to the radius of the bound state and the initial and transition temperatures of the QGP.\cite{Satz} Measurements at the CERN fixed target heavy ion program, in Pb+Pb and lighter nuclear collisions at nucleon-nucleon c.m. energy $\sqrt{s_{NN}}=17.2$ GeV,  seem to indicate ``anomalous suppression'' of the $J/\Psi$,\cite{NA50} i.e. beyond the standard nuclear absorption (the $J/\Psi$ is a hadron with $\sigma^{\rm abs}_{J/\Psi-N} \sim 6-7$mb).\cite{E772} As other models of $J/\Psi$ in a QGP indicate an enhancement due to the recombination of the free $c$ and $\bar{c}$ quarks to form quarkonia before freezeout,\cite{Thews,Kostyuk} the jury is still out, awaiting RHIC results. My summary of the different views of dilepton resonances in the High Energy\cite{UA1} and Relativistic Heavy Ion\cite{Matsui} Physics communities since the mid 1980's is shown in Fig.~\ref{fig:success}.   
\section{The Physics of Open Charm}\label{sec:2}
\subsection{Prompt Leptons}

	There are many reasons to study heavy quark production in RHI collisions: 1) production is via $g+g\rightarrow c + \bar{c}$, so charm production measures the gluon structure function and is thus sensitive to Gluon Saturation;\cite{CGC} 2) if both $J/\Psi$ and open charm are suppressed, this indicates shadowing in the gluon structure function rather than a QGP; 3) the large mass scale is very sensitive to the initial temperature; 4) the large mass scale means less radiative energy loss in a medium (i.e. QGP) compared to light quarks and detailed sensitivity to the density of color charges.\cite{deadcone} However, the key reason to study open charm and beauty is experimental: 1) the large semi-leptonic branching ratio $\sim 7-17$\% per lepton ($e$, $\mu$); 2) the large mass implies energetic leptons, $p_T\geq$ 1 GeV/c; 3) although lepton identification (i.d.) in a large hadronic and photonic background is an experimental challenge (more on this below), charm detection via a single lepton measurement has no {\bf combinatoric} background, and thus is not obviously more difficult in A+A than in p-p collisions. 

\begin{figure}[ht]
{\psfig{file=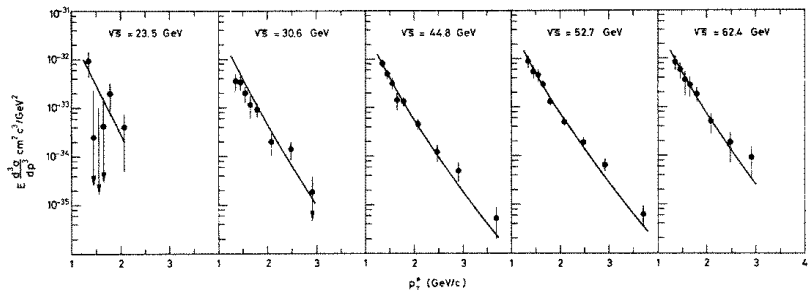,width=4.9in}}
\caption[]{Invariant cross section for $(e^+ + e^-)/2$ vs $p_T$ for 5 
values of $\sqrt{s}$ at the CERN-ISR\cite{Busser1976} compared to fits (solid lines) to corresponding data for $[(\pi^+ + \pi^-)/2] \times 10^{-4}$.  }
\label{fig:ccrs}
\end{figure}

\begin{figure}[ht]
\begin{tabular}{cc}
\centering\psfig{file=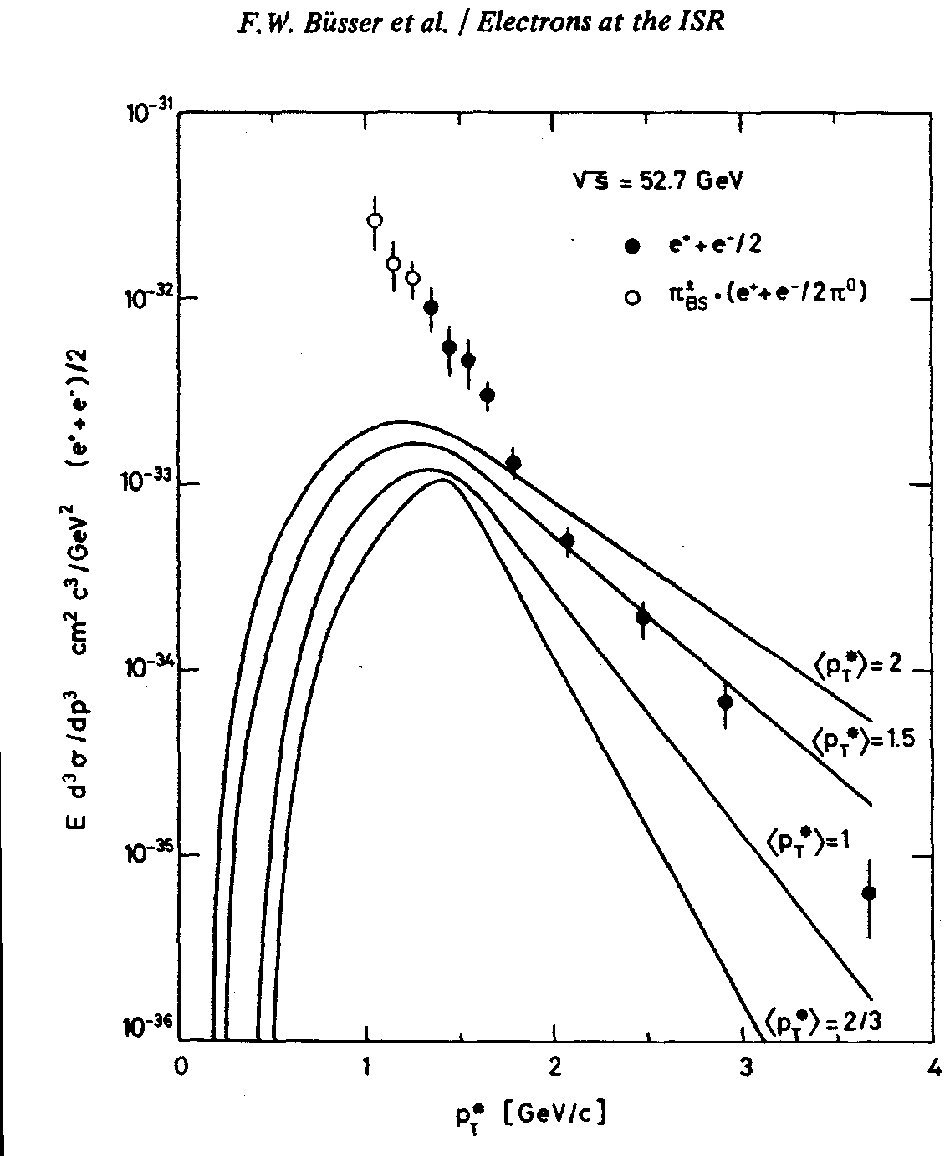,width=2.1in}
\hspace{0.014in}
\centering\psfig{file=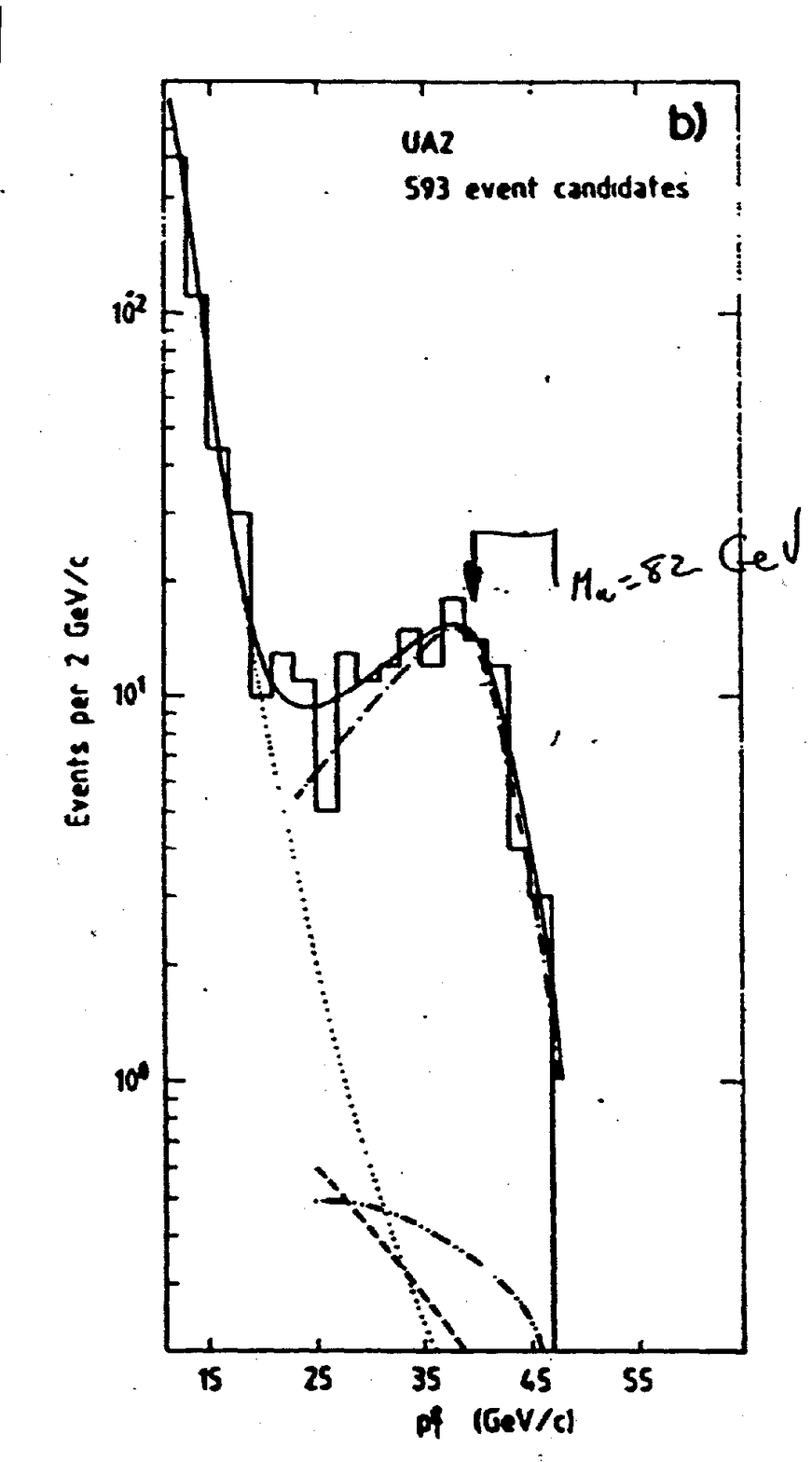,width=2.1in,height=2.5in}
\end{tabular}
\caption[]{(left) CCRS\cite{Busser1976} $e^{\pm}$ data with calculation of $e^{\pm}$ $p_T$ spectrum from $J/\Psi\rightarrow e^+ + e^-$ as a function of $\langle p_T\rangle_{J/\Psi}$ (which turned out to be $\sim 0.7-1$ GeV/c). (right) Jacobean peak from $W^{\pm}\rightarrow e^{\pm} + X$.\cite{UA2} }
\label{fig:psiW}
\end{figure}

	 It is worthwhile to remember why some experimentalists have studied high $p_T$ leptons produced in hadron collisions---they indicate Weak ($e^{\pm}\nu$) or EM ($e^+ e^-$) decays or reactions, thus possibly new physics. This idea dates back to the early 1960's when it was realized that the intermediate vector boson $W^{\pm}$ of the weak interactions\cite{LeeandYang} might most favorably be produced in nucleon-nucleon collisions.\cite{Busser1976} As the $p_T$ distribution of hadrons falls like $e^{-6 p_T}$, the lepton spectra from known hadron decays will fall faster, and can be calculated. Upon this smooth background, a heavy $W$ boson produced at rest would give a Jacobean peak at $p_{T_e}=M_{W}/2$ from the decay $W \rightarrow e +\nu$. Incredibly, this situation was described by Zichichi in 1964 in a footnote\cite{Zichichi} and was actually how the $W^{\pm}$ was discovered in 1983. However, in 1974, great excitement was generated by the discovery of prompt leptons (not from hadron decays) in p-p collisions\cite{Busser1976} at a level $e^{\pm}/\pi^{\pm}\sim 10^{-4}$ for $p_T\geq 1.3$ GeV/c, but with no Jacobean peak (see Fig.~\ref{fig:ccrs}). This was discovered before the $J/\Psi$ and turned out not to be due to the $J/\Psi$ whose Jacobean peak was well below the direct electron spectrum, in contrast to the $W^{\pm}$ discovered 9 years later (see Fig.~\ref{fig:psiW}). 
\subsection{Prompt Leptons = Charm}
	The source of the prompt leptons with $e^{\pm}/\pi^{\pm}\sim 10^{-4}$ remained a mystery for 2 years before being explained as the decay product of open charm mesons.\cite{HLBG} Meanwhile, there were possibly spurious results,\cite{Naroska} misleading conclusions\cite{Anderson} and one excellent physics idea:\cite{Glennys} the first suggestion of direct photons in p-p collisions (as the source of the prompt leptons), well before the prediction of the ``Inverse QCD compton effect''\cite{Fritzsch} $g+q\rightarrow \gamma+q$. 
See reference~\cite{PXcharm} for further discussion. 
\section{Experimental Issues}
\subsection{Real Backgrounds, Falling Spectra}
\begin{figure}[!t]
\centerline{\psfig{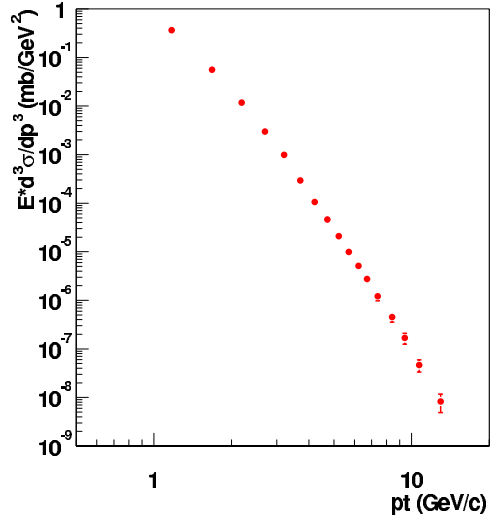}}
\vspace*{-0.25in}
\caption[]{Log-log plot of PHENIX $\pi^0$ spectrum for $\sqrt{s}=200$ GeV p-p collisions.\cite{PXQM02}}
\label{fig:pizeroll}
\end{figure}

\begin{figure}[b]
\begin{tabular}{cc}
\hspace*{-0.05in}\psfig{file=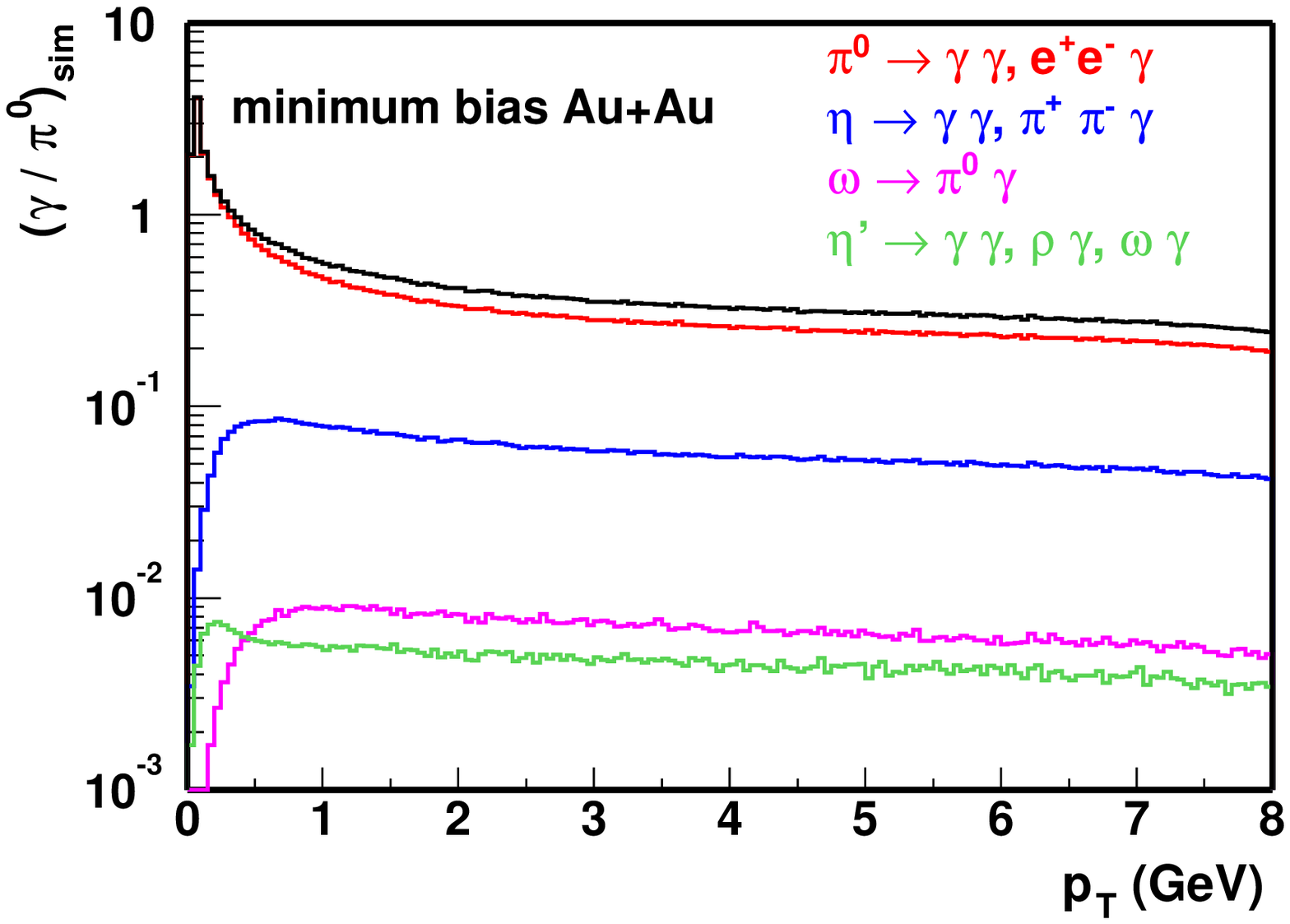,width=2.4in}&
\hspace*{-0.35in}
\psfig{file=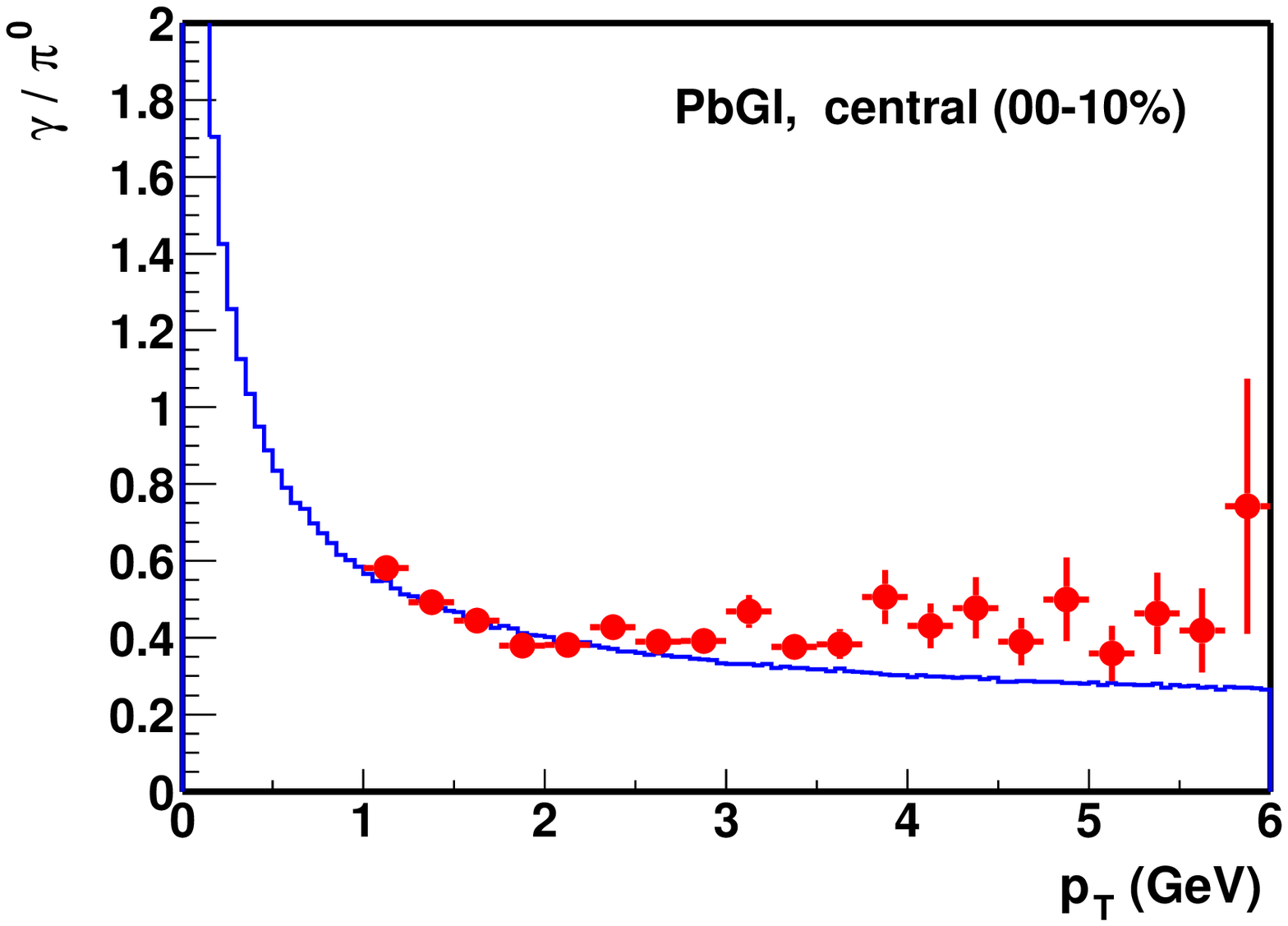,width=2.4in}
\end{tabular}
\caption[]{PHENIX\cite{PXQM02}: (left) minimum bias $\gamma /\pi^0$ from known decays;  (right) measured $\gamma /\pi^0$ from central Au+Au collisions at $\sqrt{s_{NN}}=130$ GeV (points, statistical errors only), compared to  background $\gamma/\pi^0$ from known decays (curve).}
\label{fig:pxphoton}
\end{figure}

\begin{figure}[ht]
\begin{tabular}{cc}
\hspace*{-0.05in}\psfig{file=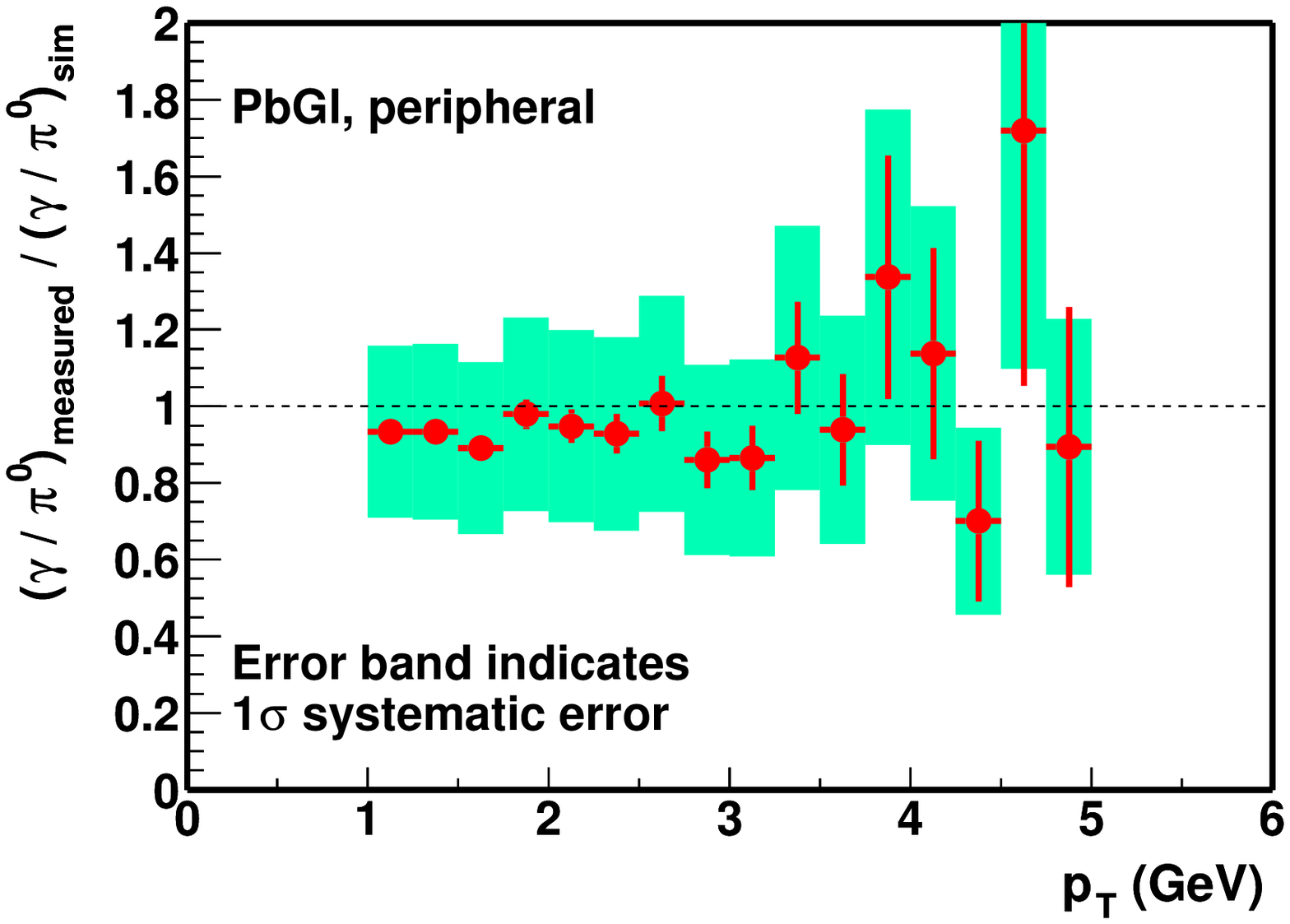,width=2.4in}&
\hspace*{-0.35in}
\psfig{file=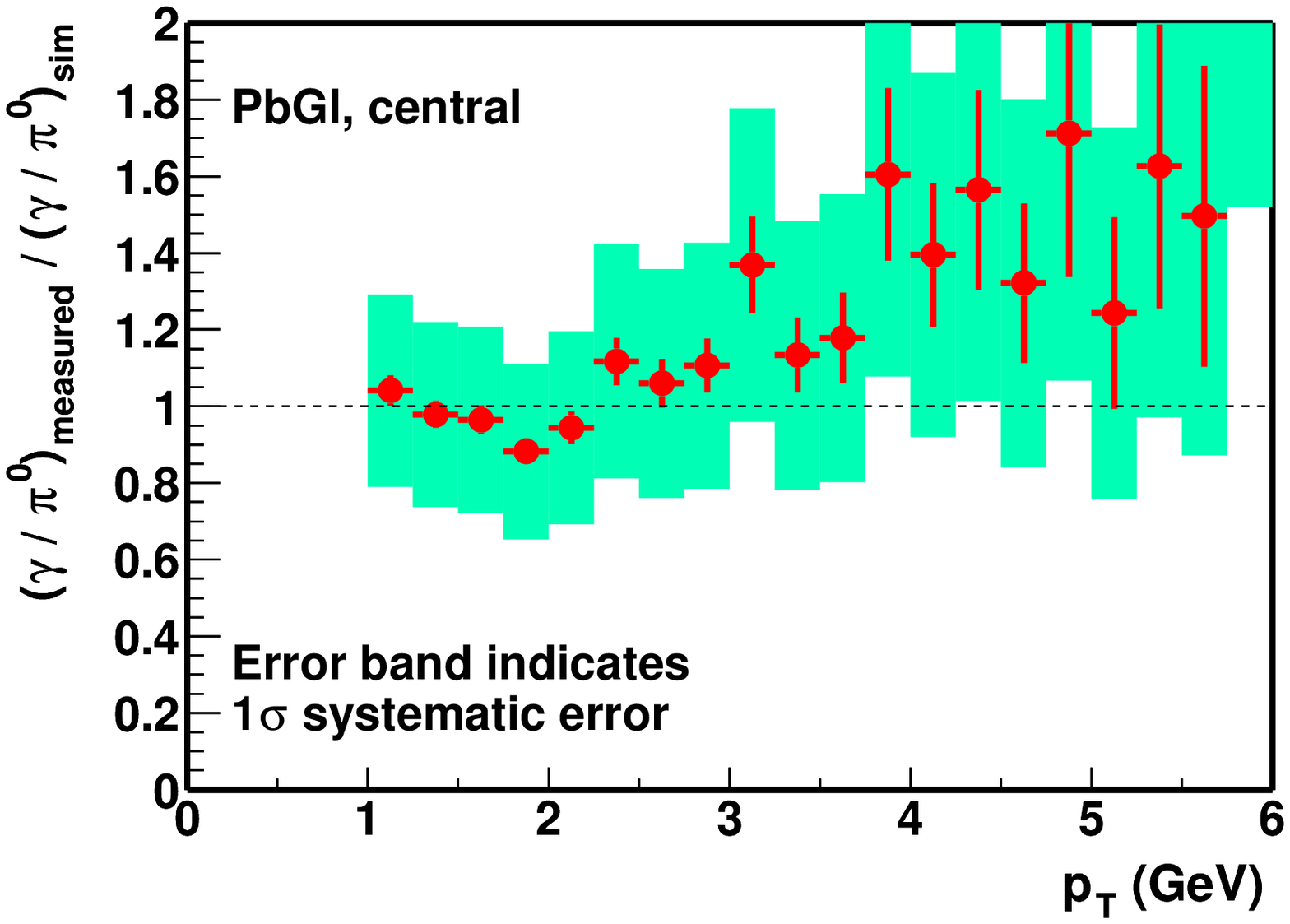,width=2.4in}
\end{tabular}
%%%%{\psfig{file=gamma_ratio_200gev.eps,width=4.7in}}

\caption[]{PHENIX\cite{PXQM02} preliminary measured $\gamma/\pi^0$ over $\gamma/\pi^0$ from known decays in 200 GeV Au+Au collisions. The shaded boxes represent the estimated $1\sigma$ systematic errors.}
\label{fig:pxprelimgamma}
\end{figure}

   The problem with photon and lepton ($e,\mu$) searches in p-p and A+A collisions is that there is a huge background of real photons from $\pi^0\rightarrow \gamma + \gamma$, $\eta \rightarrow \gamma +\gamma$ and other decays. This background can be calculated once the $\pi^0$ and $\eta$ $p_T$ spectra are known. In the region of high $p_T\geq 3$ Gev/c where the QCD photons\cite{Fritzsch} are expected the spectra are excellent power laws (see Fig.~\ref{fig:pizeroll})\cite{PXQM02} and it is easy to show that if:
\begin{equation}
{{dn_{\pi^0}}\over{p_T dp_T}}\propto p_T^{-n}\qquad \mbox{then}\qquad 
\left . {\gamma\over \pi^{0}}\right |_{\pi^0}\!\! (p_T)=2/(n-1) \qquad. 
\label{eq:powerlaw}
\end{equation}
In Fig.~\ref{fig:pizeroll}, $n\sim 8$, so $\gamma|_{\pi^0}/\pi^0\sim 1.2\times2/7=0.34$ where  the factor 1.2 includes $\eta\rightarrow \gamma+\gamma$ (estimated). The validity of this simple approach is evident from the full calculation of $\gamma/\pi^0$ from known decays for PHENIX 200 GeV Au+Au $\pi^0$ data\cite{PXQM02} compared to the measured semi-inclusive $\gamma/\pi^0$ ratio (Fig.~\ref{fig:pxphoton}).  
Only statistical errors are shown. A key systematic uncertainy for photon measurements is the possibile non-linearity of the photon energy measurement (in an ElectroMagnetic Calorimeter): Eq.~\ref{eq:powerlaw} assumes that e.g. one 10 GeV photon and two 5 GeV photons (a 10 GeV $\pi^0$) measure at exactly the same energy in the experiment. The preliminary PHENIX result with systematic errors included is inconclusive (Fig.~\ref{fig:pxprelimgamma}). 

STAR\cite{STQM02} uses a different method of photon detection than PHENIX, measurement of converted photons. This avoids the non-linearity problems of EMCalorimeters, and measures the photon direction, but suffers from the low-probability-squared of converting both photons from a $\pi^0$, which causes a large systematic uncertainty due to lack of knowledge of the detailed shape of the $\pi^0$ spectrum. If STAR uses the PHENIX $\pi^0$ spectrum, then the issue of the relative and absolute accuracy of the $p_T$ scales comes into play, which is clearly important in spectra which fall like the $n\sim 8^{\rm th}$  power. 
\subsection{Detecting Electrons and Photons}

\begin{figure}[ht]
\begin{tabular}{cc}
\centering\psfig{file=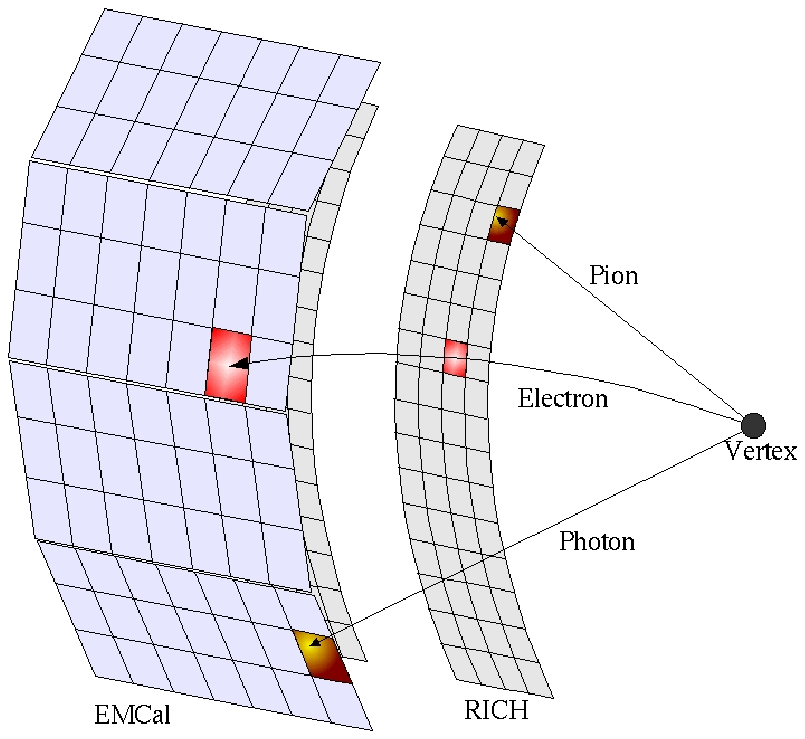,width=2.4in}
\hspace{0.014in}
\centering\psfig{file=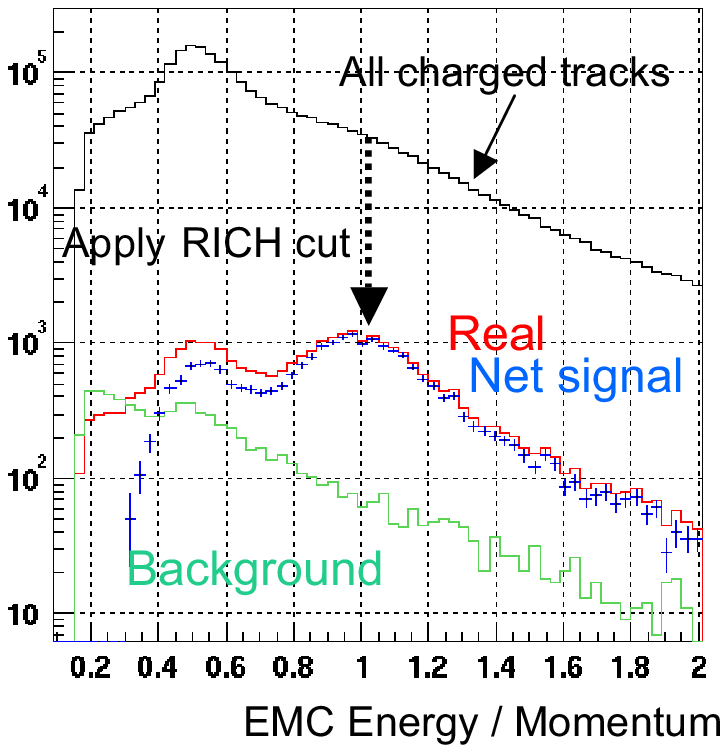,width=2.0in}
\end{tabular}
\caption[]{Schematic of $\pi^{\pm}$, $e^{\pm}$ and $\gamma$ in PHENIX, with ElectroMagnetic Calorimter (EMCal) and Ring Imaging Cherenkov Counter (RICH). (right) Energy/momentum for all charged particles detected in the EMcal with and without a RICH signal.}
\label{fig:pxworks}
\end{figure}

	Electron and photon detection are intimately connected and require an open geometry, which also allows a hadron measurement. Muons are identified by passage through a thick absorber, which, in general, precludes measurement of any other particles. A schematic drawing of the PHENIX electron/photon detector\cite{KBarish} is shown in Fig.~\ref{fig:pxworks}.   
The EMCal measures the energy of $\gamma$ and $e^{\pm}$ and reconstructs $\pi^0$ from 2 photons. It measures a decent time of flight (TOF), 0.3 nanoseconds over 5 meters, allowing photon and charged particle identification. Electrons are identified by a count in the RICH (cherenkov) and matching Energy and momentum ($E/p$), where the momentum is measured by track chambers in a magnetic field. Charged hadrons deposit only minimum ionization in the EMCal ($\sim 0.3$ GeV), or higher if they interact, and don't count in the RICH ($\pi^{\pm}$ threshold 4.7 GeV/c). Thus, requiring a RICH signal rejects all charged hadrons with $p < 4.7$ GeV/c, leaving only $e^{\pm}$ as indicated by the $E/p=1$ peak in Fig.~\ref{fig:pxworks}(right). A high precision TOF over part of the aperture allows improved charged hadron id.\cite{JT} It is amusing  to realize that once you decide to measure electrons, you must make an excellent $\pi^0$ measurement to understand the background, and this implies a detector which can measure and identify almost all particles. 
\subsection{All Photons Create Background $e^+ e^-$ Pairs}
\begin{figure}[t]
\begin{tabular}{cc}
\hspace*{-0.12in}
{\psfig{file=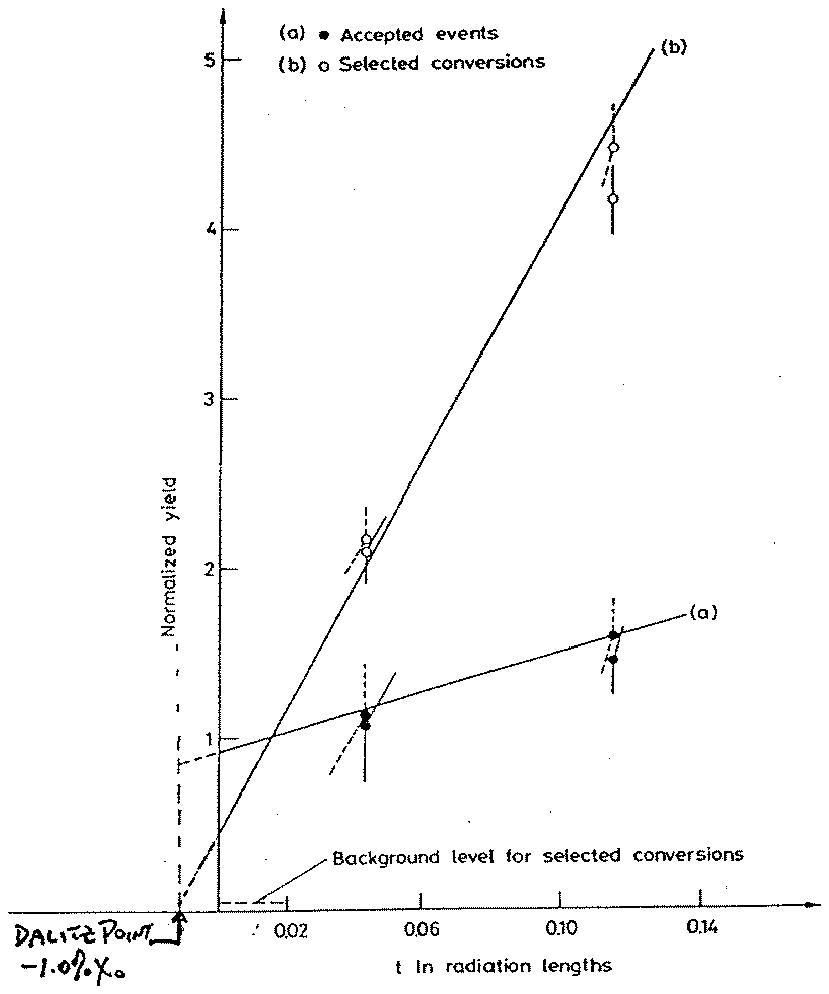,width=2.2in}}&
\hspace*{-0.12in}\psfig{file=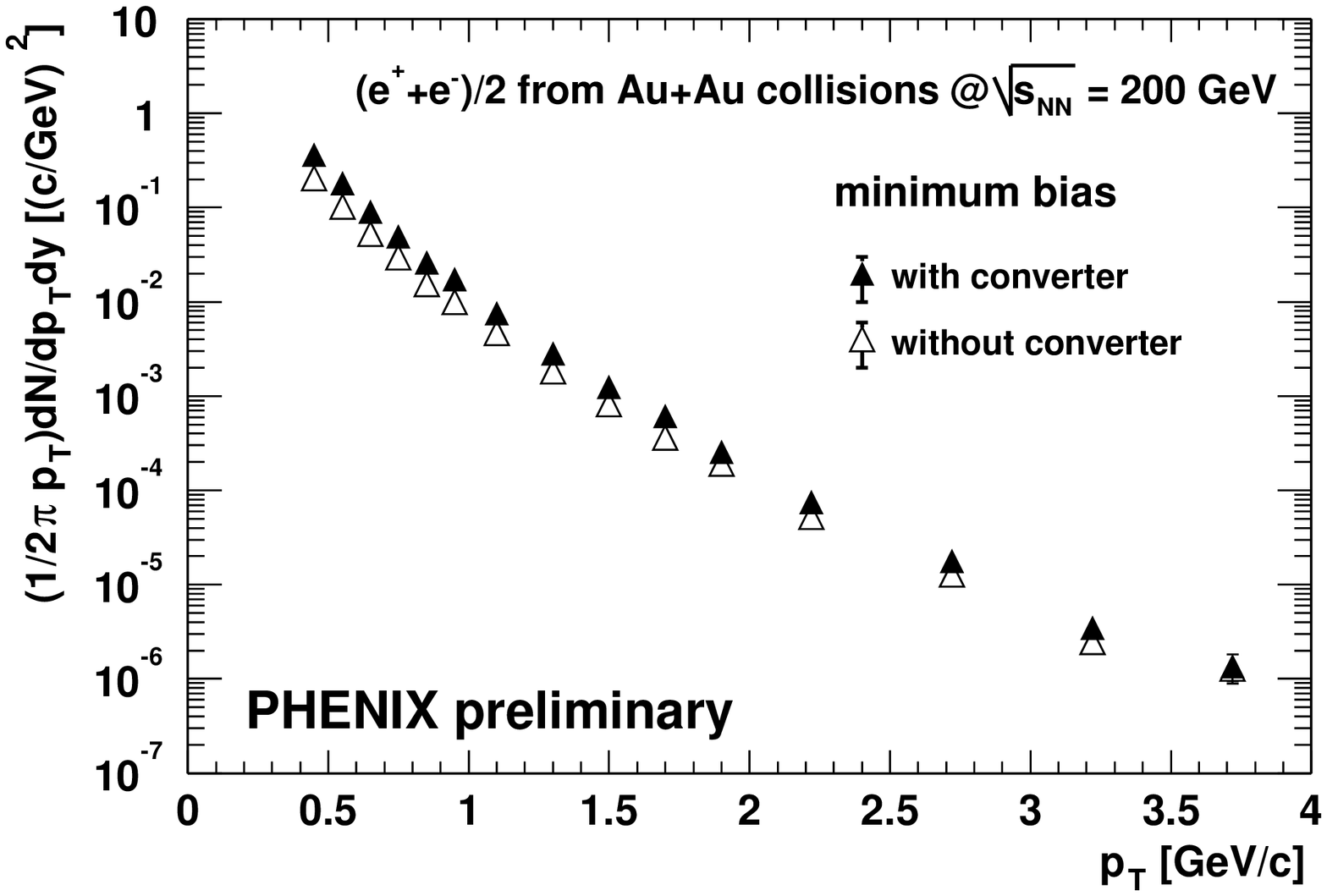,width=2.2in,height=2.4in}\vspace*{0.05in}
\end{tabular}
\caption[]{({left}) CCRS\cite{Busser1976} yield of inclusive electrons vs total external radiation length $t/X_0$ for (b) selected conversions (extrapolates to zero at the ``Dalitz point") and (a) Accepted events with prompt electrons (non-zero intercept at Dalitz point). Yields are relative, normalized to 1 at the standard thickness $t/X_0=0.016$. ({right}) PHENIX\cite{PXe} $e^{\pm}$ yield with and without an external converter.   }
\label{fig:xrad}
\end{figure}

All photons create background $e^+ e^-$ pairs by external or internal conversion. This allows a precision ($ < 1$\%) cross calibration of $e^{\pm}$ and $\gamma$ energies, as in Fig.~\ref{fig:pxworks}(right), eliminating any non-linearity problem as in the $\gamma$-$\pi^0$ energy comparison. The probability of internal and external conversion per $\gamma$ is 
\begin{equation}
 {e^{-}|_{\gamma} \over \gamma} = {e^{+}|_{\gamma} \over \gamma} 
={ \delta_{2} \over 2} + {t \over { {9\over 7} X_0} } \equiv \delta_{eff} 
\label{eq:efromgamma}
\end{equation}  
where $\delta_{2}/2=$ Dalitz (internal conversion) branching ratio per photon 
= 0.6\% for $\pi^0\rightarrow \gamma \gamma$, 0.8\% for $\eta\rightarrow \gamma \gamma$.\footnote{Note the insensitivity to the $\eta/\pi^0$ ratio.} Clearly, the external $t/{9\over 7}X_{0}$ must be comparable 
$\sim 0.6$\% to avoid too much additional background from external conversions. This sets a very severe radiation length budget for an $e^{\pm}$ detector. However, one can add small external converters of a few \% $X_0$ in a test run (see Fig.~\ref{fig:xrad}) to determine whether, as for a pure photonic source (Eq.~\ref{eq:efromgamma}), $(e^{+}+e^{-})/2\gamma\rightarrow 0$ at the ``Dalitz Point",  $t/X_{0}=-{9\over 7}\delta_{2}/2\sim 0.8-1.0$\% in units of radiation lengths.
Also, note that in Fig.~\ref{fig:xrad}(left), the photonic curve (b) increases much more rapidly with added converter than the prompt $e^{\pm}$ candidate  curve (a). In the PHENIX\cite{PXe} converter measurement Fig.~\ref{fig:xrad}(right), the lower $p_T$ points show a much larger converter effect than the higher $p_T$ points indicating a clear non-photonic component at higher $p_T$. 
\subsubsection{Effect of the Falling Spectrum}
\begin{figure}[ht]
\begin{tabular}{cc}
\centering\psfig{file=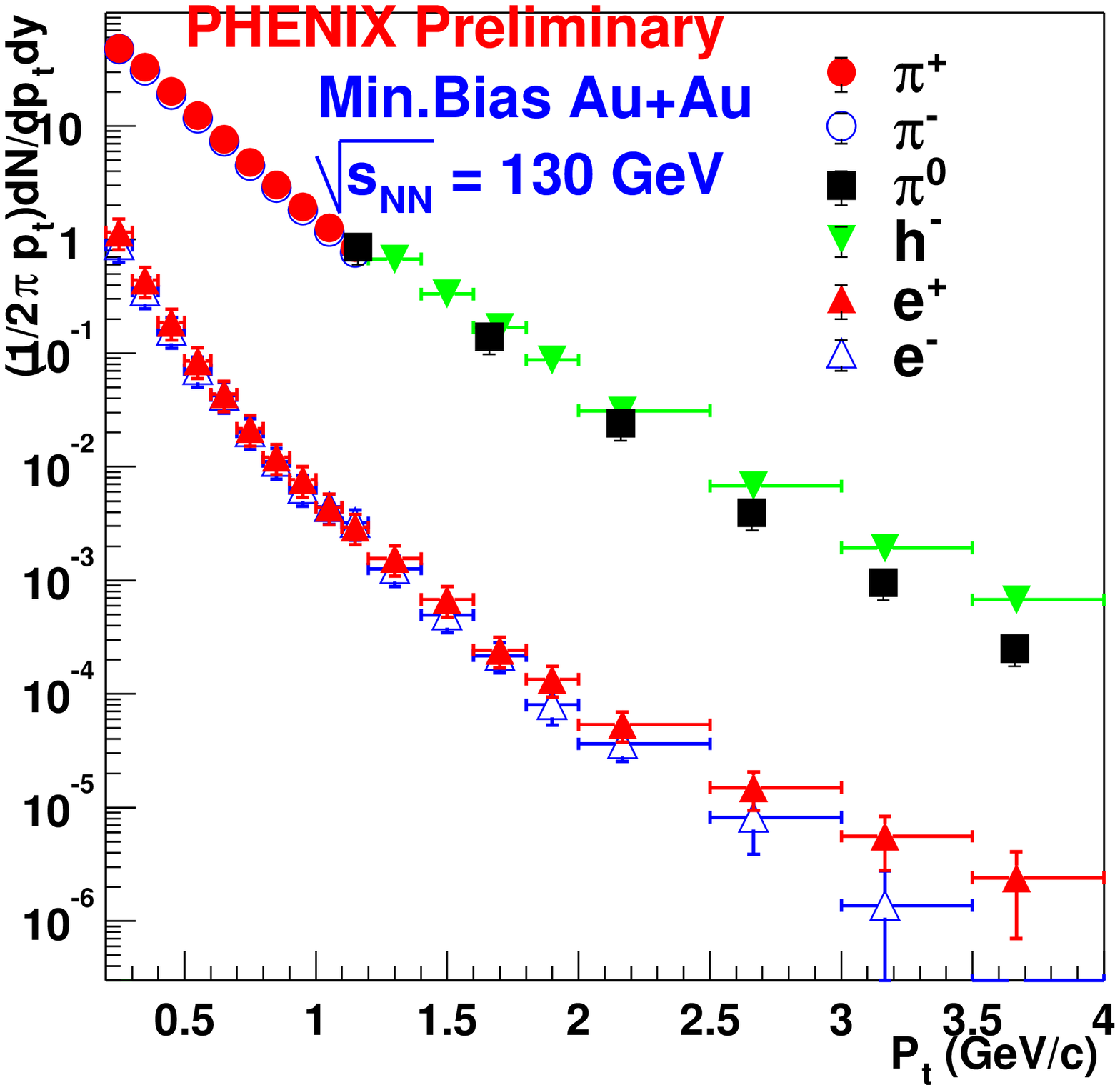,width=2.4in}
&\hspace*{-0.02in}
\centering\psfig{file=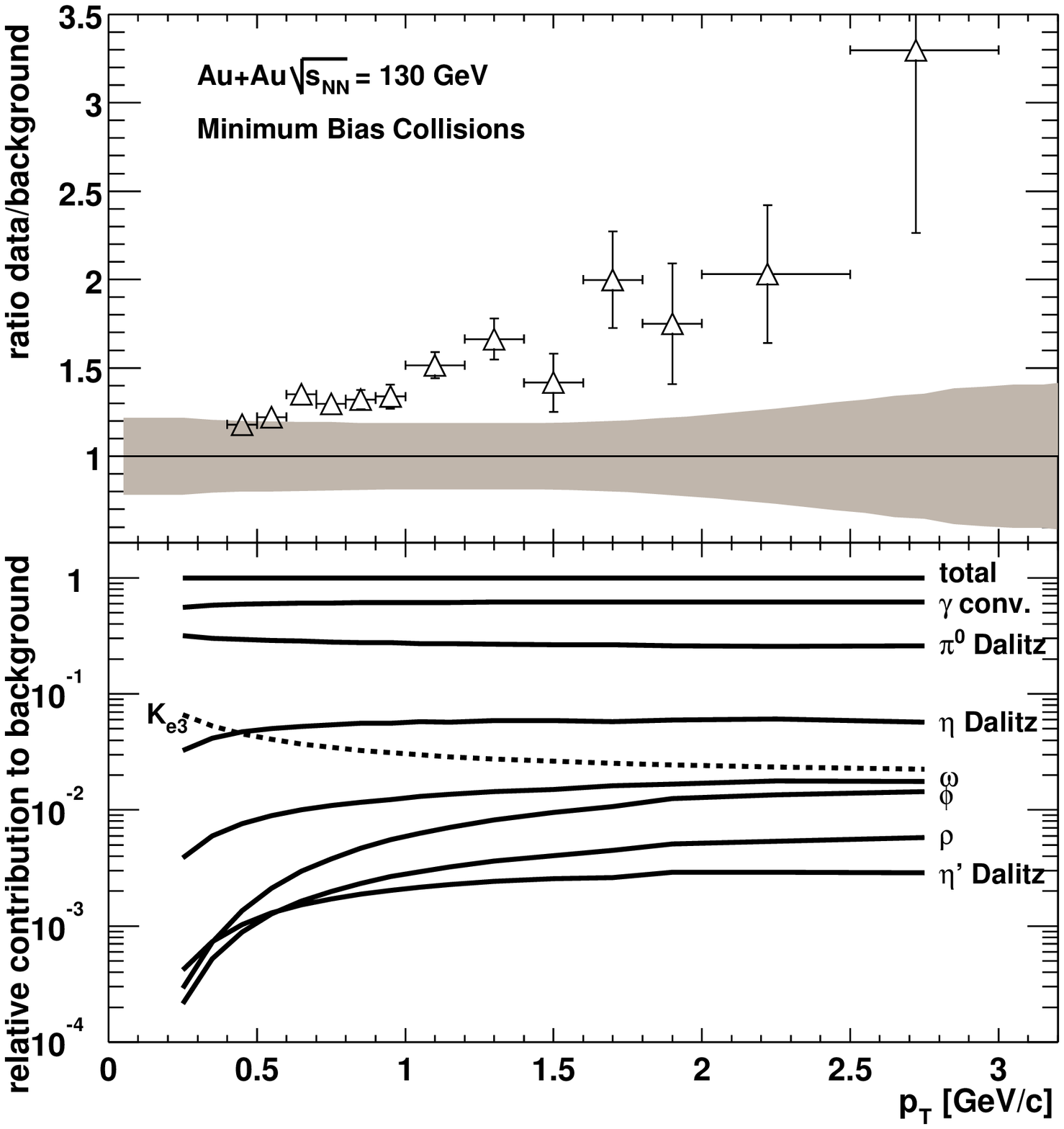,width=2.2in}
\end{tabular}
\caption[]{PHENIX:\cite{PXe} (left) Measured inclusive $p_T$ spectra for identified $\pi^{\pm}$, $\pi^0$, $e^{\pm}$ and non-identified $h^{\pm}$ charged particles in Au+Au collisions at $\sqrt{s_{NN}}=130$ GeV; (right) Measured $(e^{+}+e^{-})/2\pi^0$ divided by the expected background. The shaded region is the systematic error. The relative components of the background are also shown.}
\label{fig:pxepi}
\end{figure}

    Still using the $p_T^{-n}$ power law for the $\pi^0$ and thus the decay $\gamma$ spectra (Eq.~\ref{eq:powerlaw}) one finds: 
\begin{equation} 
 \left . {e^{-}\over \pi^{0}}\right |_{\pi^0}\!\! (p_T)= 
\left . {(e^{-} + e^{+})\over 2\pi^{0}}\right |_{\pi^0}\!\! (p_T) = 
\delta_{eff}\times {2\over {(n-1)^2}}
\label{eq:eoverpi}
\end{equation}
which for $n=8$ gives $e^{-}|_{\pi^0}/\pi^0 > 0.6\% /7^{2}=1.2\times 10^{-4}$. Thus one needs $\sim 10^{4}$ rejection against $\pi^{\pm}$ 
just to be able to see the $e^{\pm}$ background  from $\pi^0$ Dalitz. The RICH  and EMCal (Fig.~\ref{fig:pxworks}) give $>10^5$ rejection, but for the record, the measured $e/\pi$ is large in Au+Au minimum bias collisions at RHIC,\cite{PXe} $\sim 1/500=2\times 10^{-3}\gg 1.2\times 10^{-4}$ (see Fig.~\ref{fig:pxepi}). 
Note that the conversion/dalitz background is decent, only $1.8\pm 0.2$. Thus, since the measured $e/\pi$ is significantly larger than that expected from background sources, beyond the systematic error for $p_T >0.6$ GeV/c, a clear prompt $e^{\pm}$ signal is observed, the first measurement of charm in A+A collisions.  
\section{Charm and $J/\Psi$ Experimental Results from RHIC}
\begin{figure}[ht]
\begin{tabular}{cc}
\centering\psfig{file=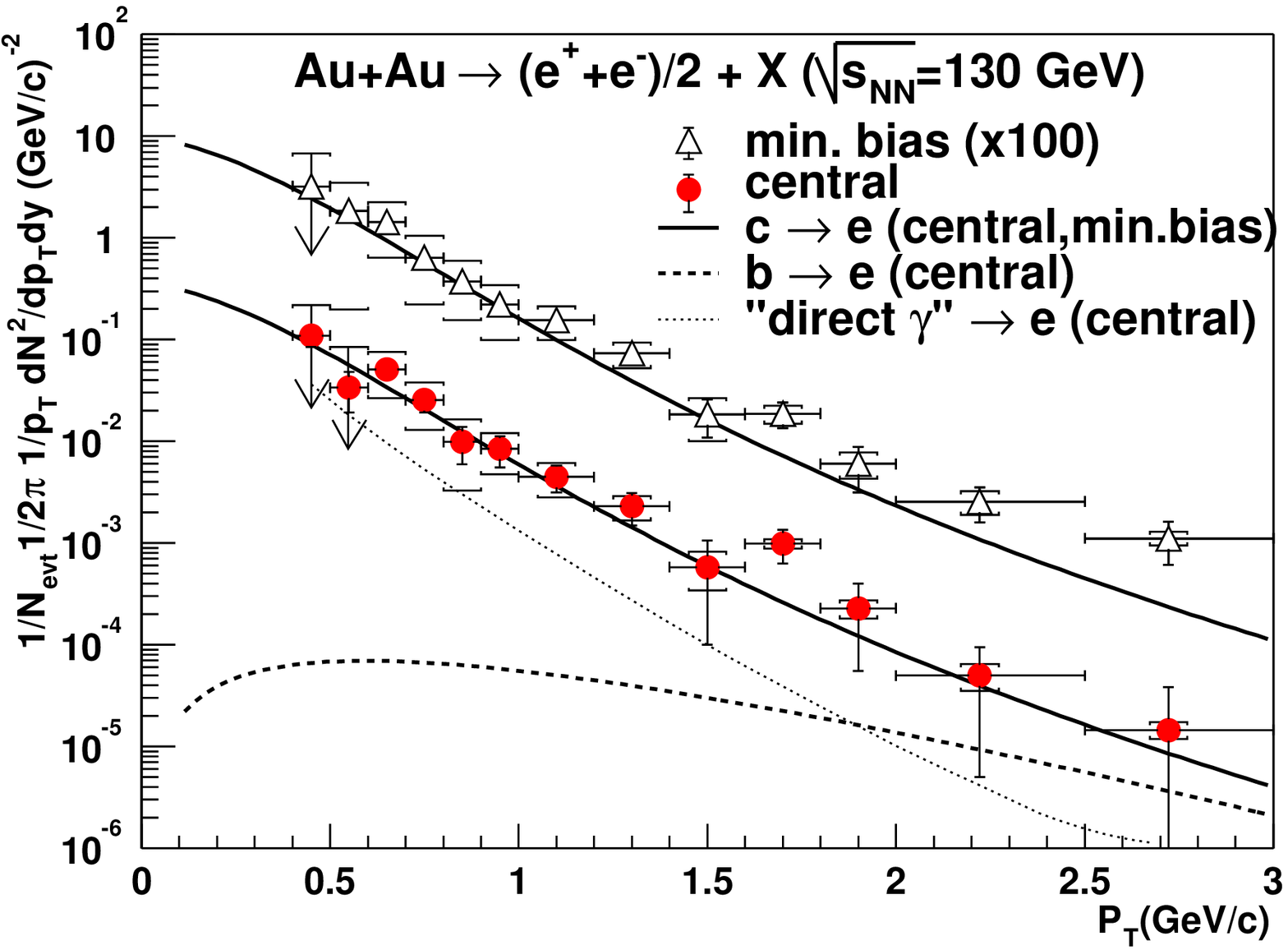,width=2.2in}
\hspace{0.014in}
\centering\psfig{file=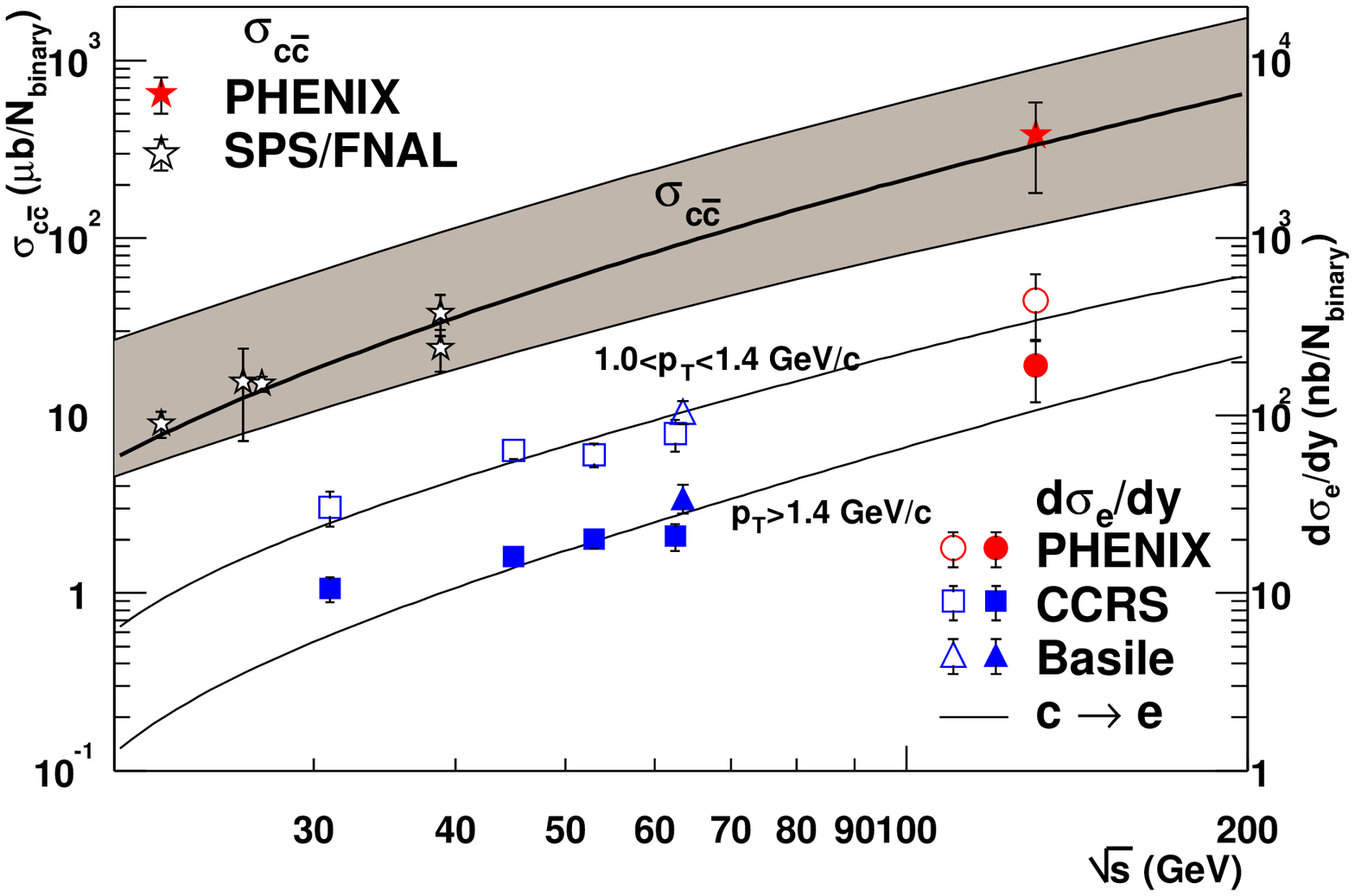,width=2.2in}
\end{tabular}
\caption[]{PHENIX measurements:\cite{PXe} (left) Background subtracted electron $p_T$ spectra for minimum bias (scaled up by a factor of 100) and central Au+Au collisions compared with the expected\cite{PYTHIA} contributions from open charm, and, for central collisions, from beauty and internal conversion of QCD direct $\gamma$. (right) Single electron cross sections $d\sigma_{e}/dy|_{y=0}$ per N-N collision, compared to lower energy p-p measurements in two $p_T$ ranges, along with open charm calculations from PYTHIA (bottom, right-hand scale). The derived charm cross section $\sigma_{c\bar{c}}$ compared to lower energy measurements (top, left-hand scale)---the thick curve and the shaded band are PYTHIA\cite{PYTHIA} and NLO pQCD\cite{NLOcharm} predictions, respectively.}
\label{fig:pxcharm}
\end{figure}

%\subsection{Charm} 
The background subtracted electron spectra\cite{PXe} from  Fig.~\ref{fig:pxepi} are shown in Fig.~\ref{fig:pxcharm}(left) for minimum bias and central 
collisions (data points) together with the expected contributions from open charm\cite{PYTHIA} and beauty decays (lines) assuming point-like (binary-collision) scaling of the charm cross section which agrees very well with the measurements. Thus charm does not appear to be suppressed compared to point-like expectations, in sharp contrast to the $\pi^0$ which are suppressed\cite{PXQM02}  by a factor of 3--5! The measured cross section  $d\sigma_{e}/dy|_{y=0}$ per N-N  collision and the derived ($c\bar{c}$) total cross section are extracted assuming point-like scaling and compared to p-p data from lower $\sqrt{s}$ and an NLO pQCD calculation\cite{NLOcharm}, which agree very well Fig.~\ref{fig:pxcharm}(right).     
%\subsection{$J/\Psi$}
 
\begin{figure}[!hb]
\begin{center}
\begin{tabular}{cc}
\centering\psfig{file=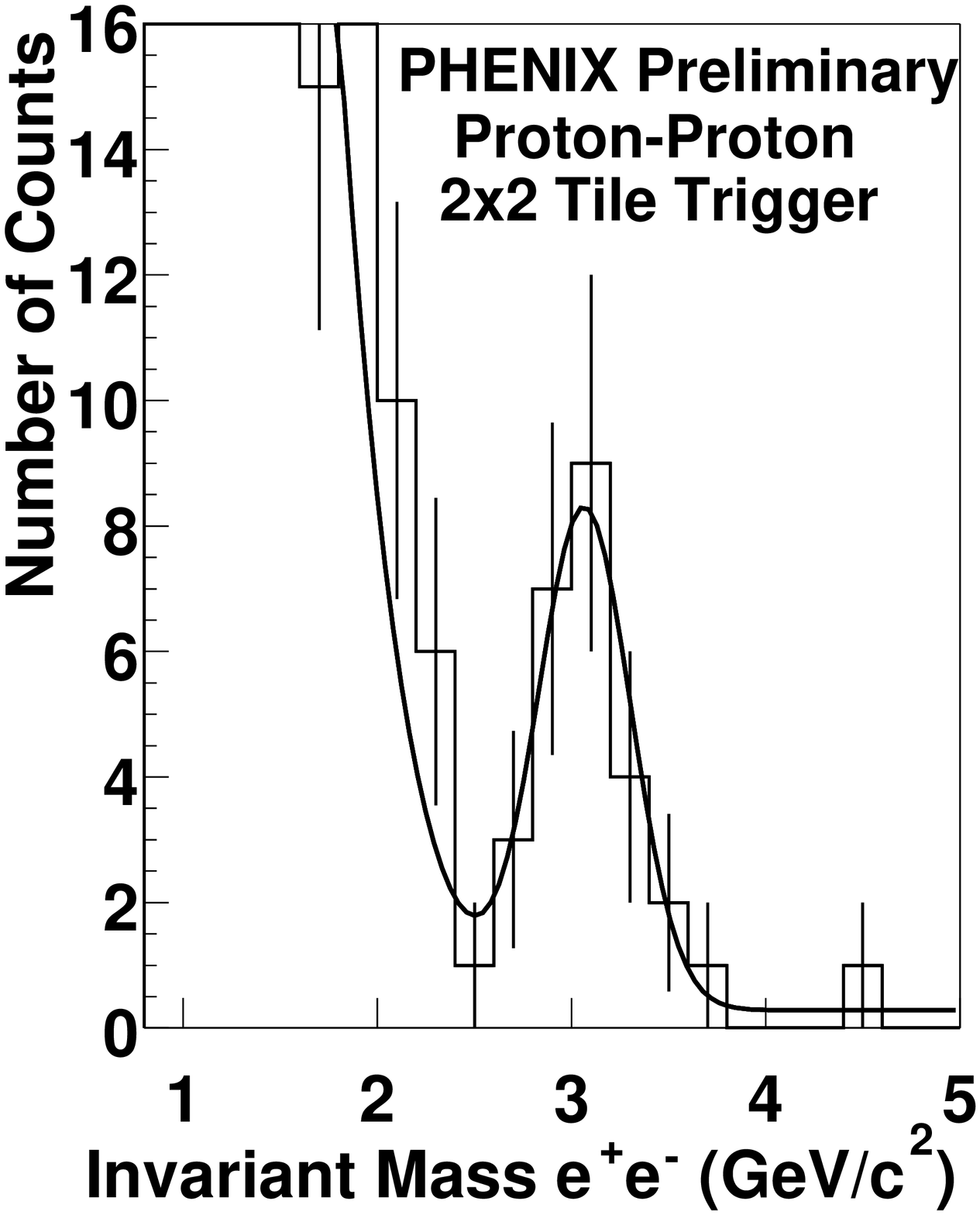,width=1.8in}&
\hspace{0.014in}
\centering\psfig{file=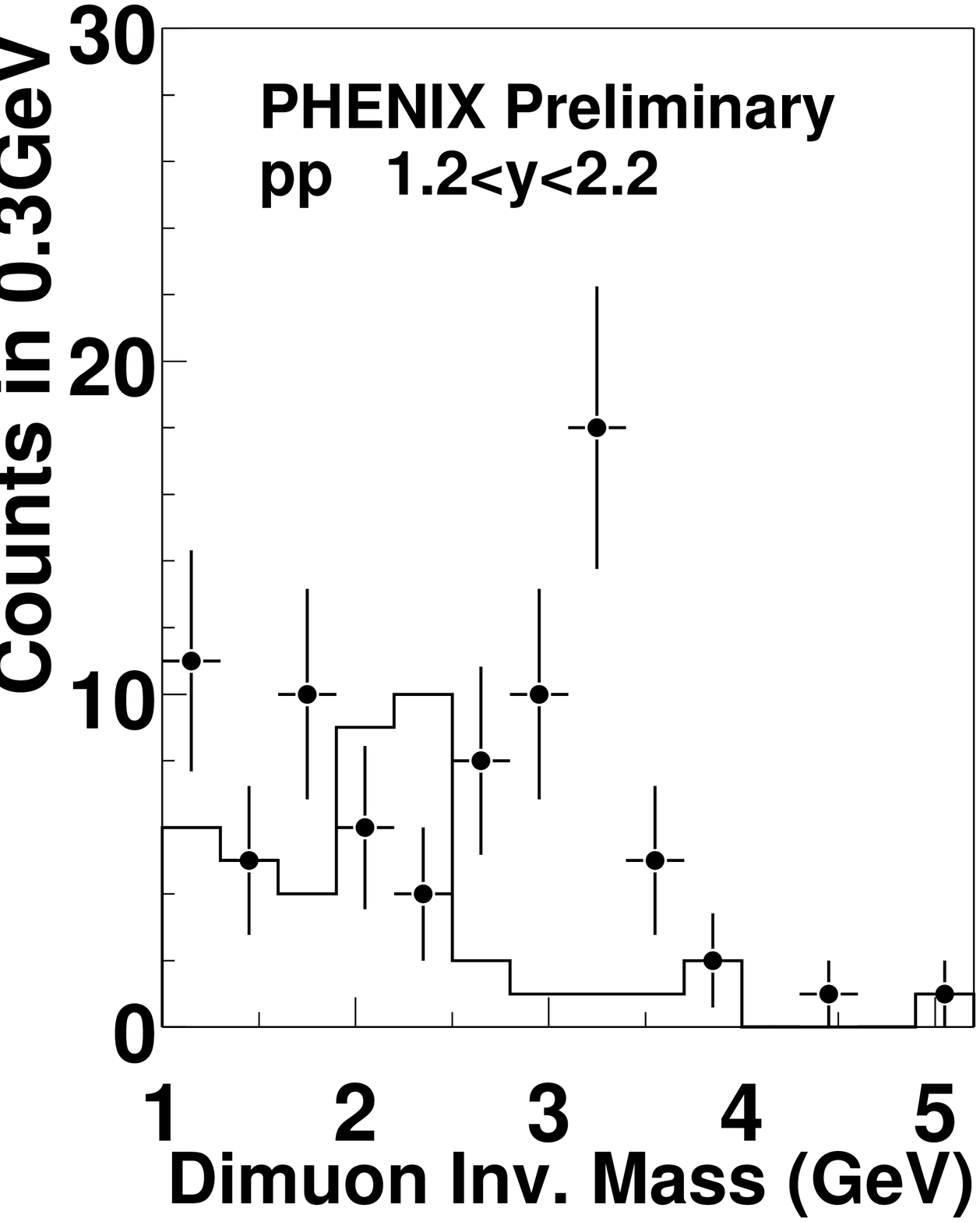,width=1.8in}
\end{tabular}
\end{center}
\vspace*{-0.12in}
\caption[]{PHENIX:\cite{PXe} (left) $p+p\rightarrow e^+ + e^- +X$ invariant mass spectrum for $|y|\leq 0.35$; (right) $p+p\rightarrow \mu^+ + \mu^- +X$ invariant mass spectrum for $1.2 < y <2.2$.}
\label{fig:ppjpsimass}
\vspace*{-0.12in}
\end{figure}

\begin{figure}[!t]
\begin{tabular}{cc}
\centering\psfig{file=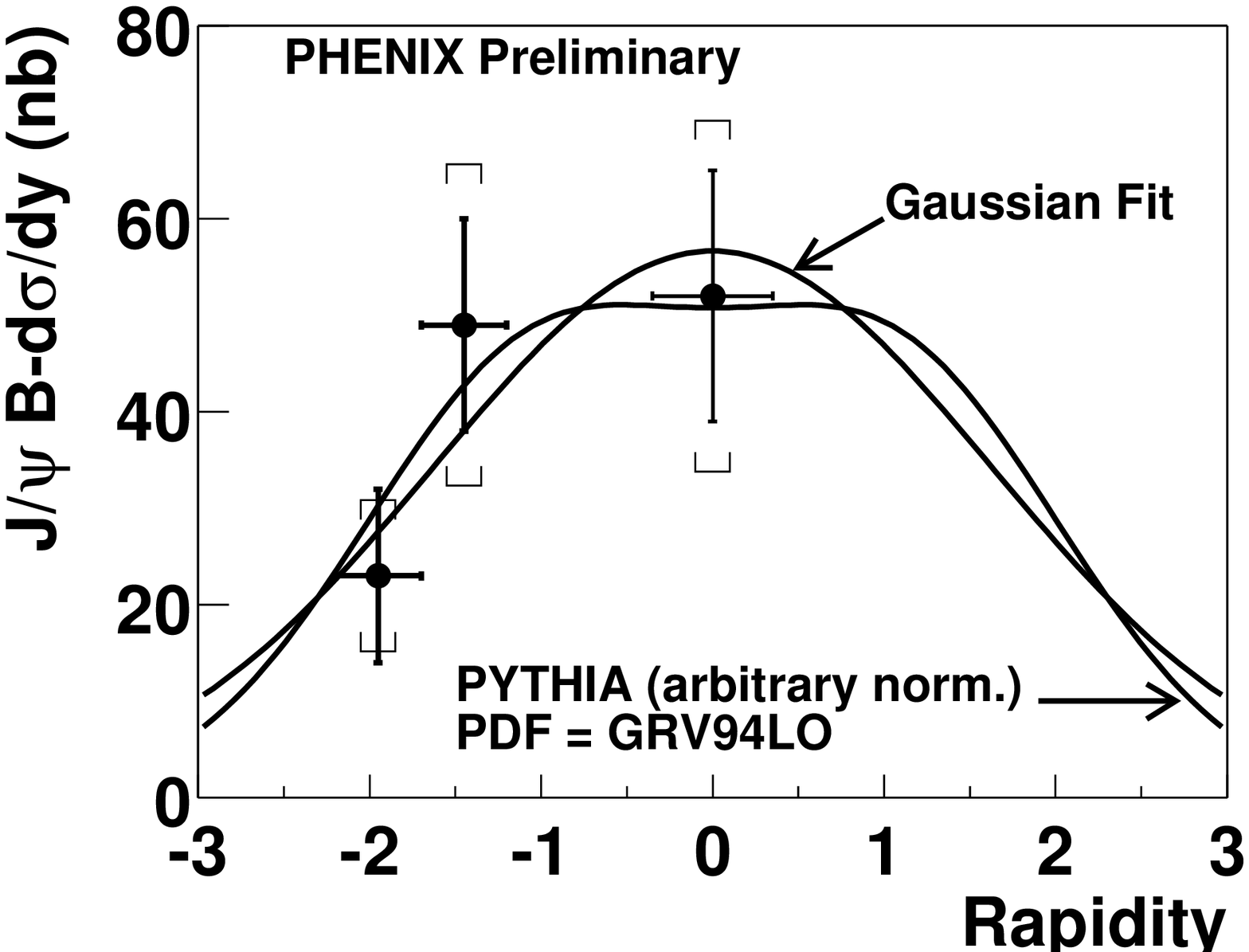,width=2.2in}
\hspace{0.014in}
\centering\psfig{file=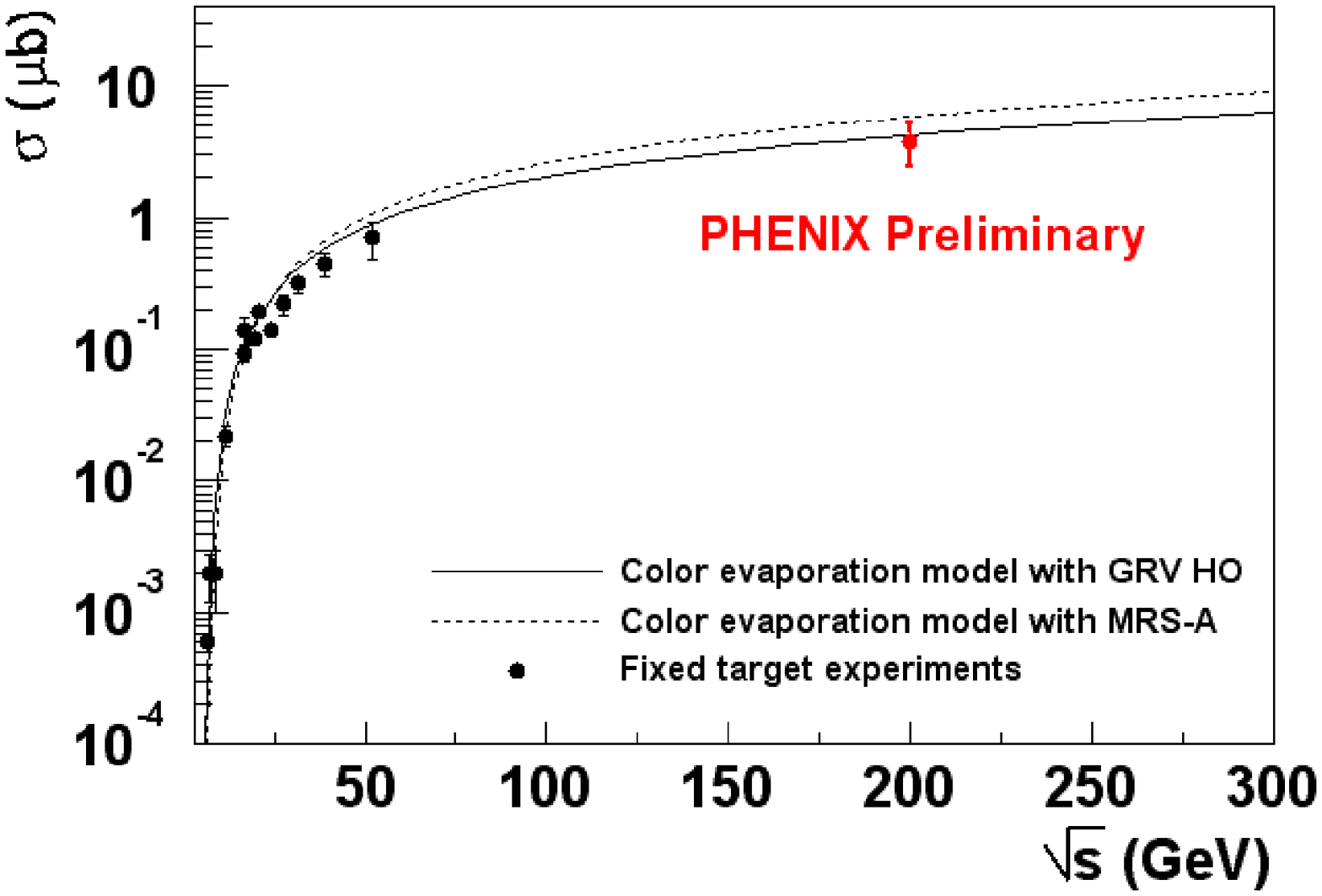,width=2.2in}
\end{tabular}
\caption[]{(left) $B_{ll} d\sigma/dy$ as a function of rapidity for $pp\rightarrow J/\Psi (\rightarrow l^+ l^{-}) + X$. Brackets represent the systematic errors. (right) $J/\Psi$ cross section integrated over rapidity compared to measurements at lower $\sqrt{s}$ and color evaporation model.}
\label{fig:sigjpsipp}
\end{figure}

    The first preliminary measurements of $J/\Psi$ production at RHIC have been obtained by the PHENIX collaboration\cite{PXe} in p-p and Au+Au collisions at $\sqrt{s_{NN}}=200$ GeV. The p-p data are shown in Fig.~\ref{fig:ppjpsimass}.  
A total of 24 $J/\Psi\rightarrow e^+ +e^-$ and 26 $J/\Psi\rightarrow \mu^{+}  + \mu^{-}$ events are observed, a far cry from the state of the art in Fig.~\ref{fig:cfs}. Nevertheless, these data measure $B d\sigma/dy$ over a large range of rapidity (see Fig.~\ref{fig:sigjpsipp}) so that the total 
$J/\Psi$ cross section can be measured using a Gaussian fit to the data, which agrees with the PYTHIA\cite{PYTHIA} prediction for the rapidity distribution. The measured cross section compares favorably with measurements\cite{Schub} at lower $\sqrt{s}$ and follows the trend predicted by the Color Evaporation Model.\cite{CEM}

\begin{figure}[htb]
\begin{tabular}{cc}
\centering\psfig{file=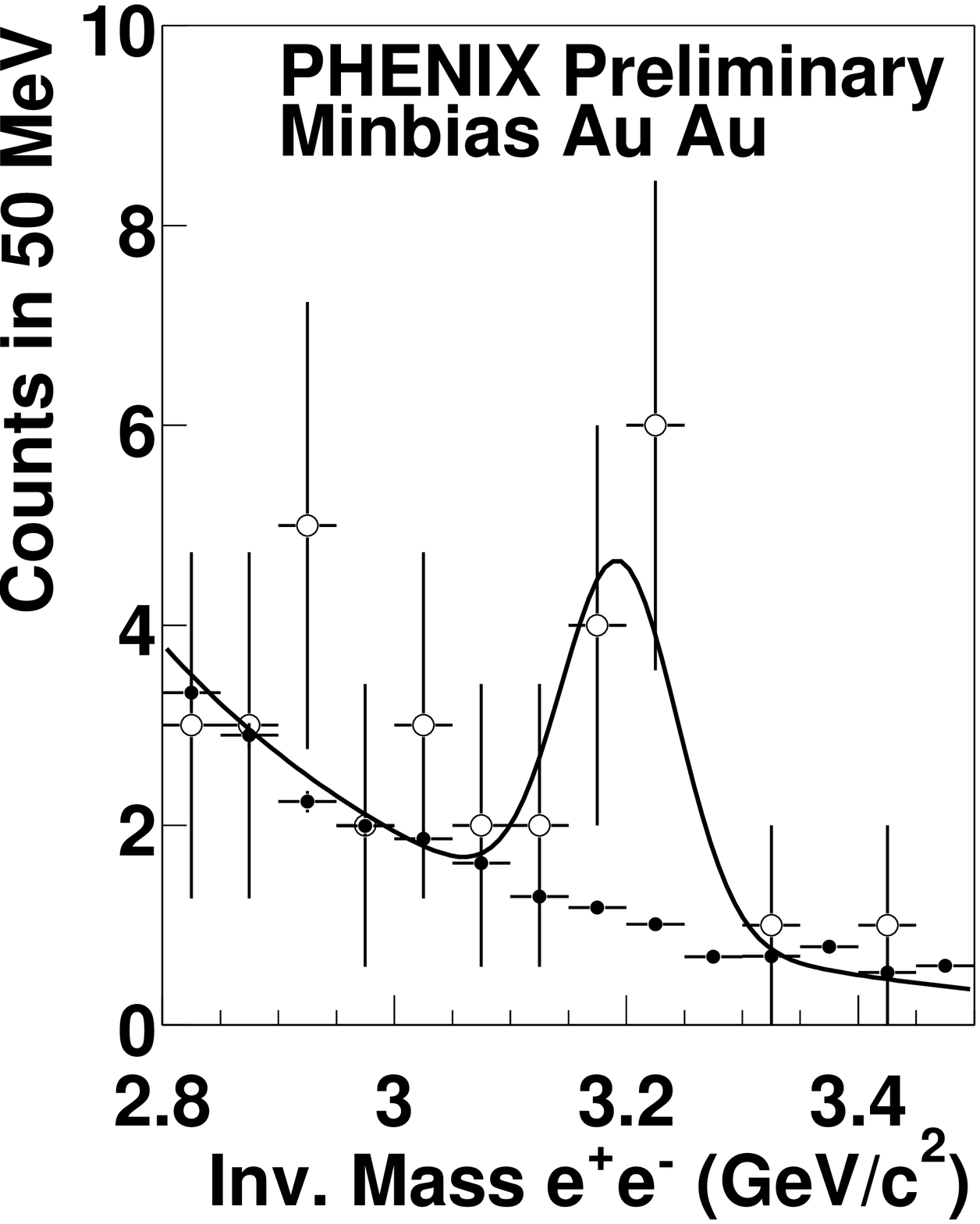,width=1.7in}
\hspace{0.014in}
\centering\psfig{file=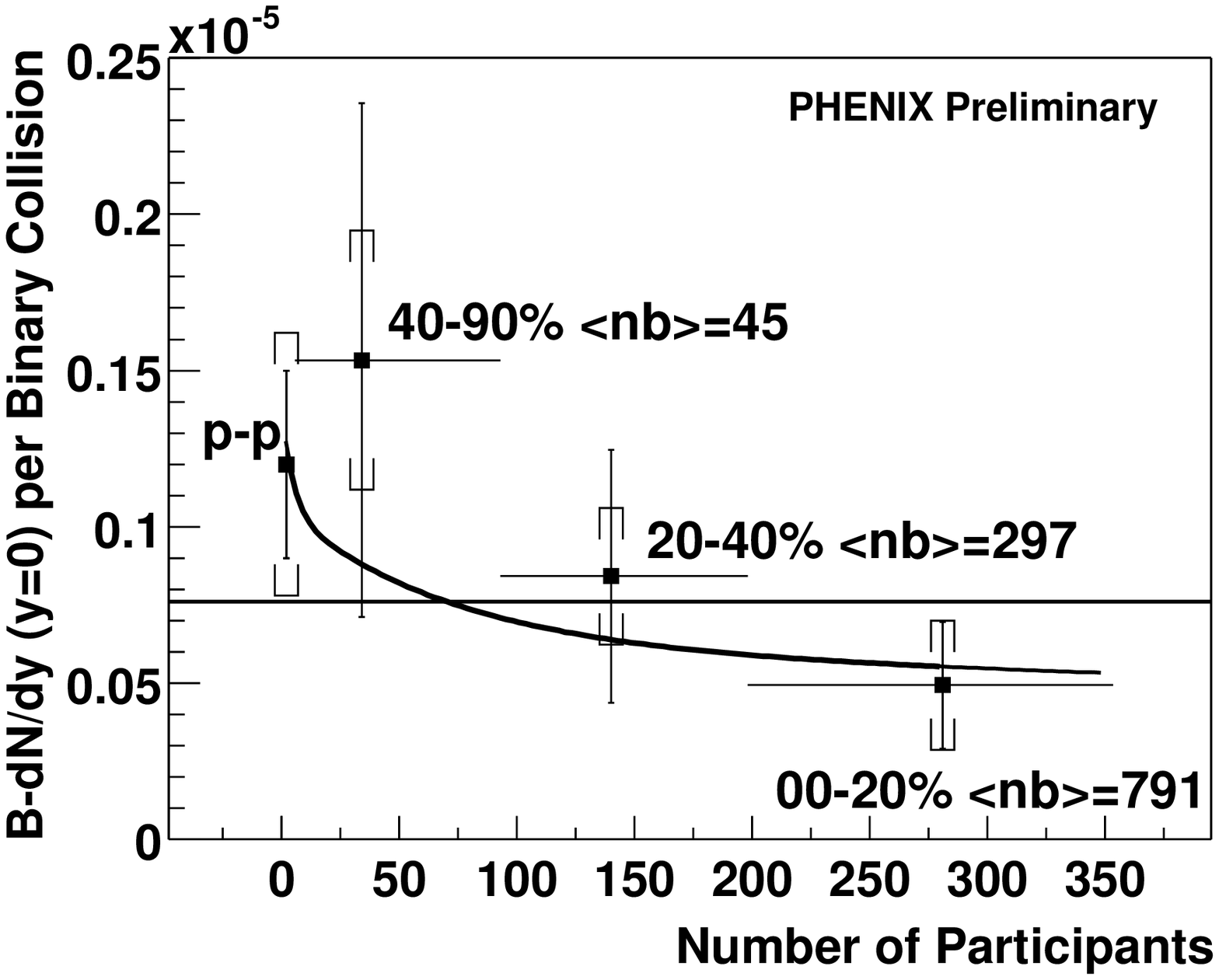,width=2.8in}
\end{tabular}
\caption[]{(left) PHENIX $e^+ e^-$ invariant mass spectrum for $|y|\leq 0.35$ in minimum bias Au+Au collisions, together with the mixed event background distribution. (right) $B_{ee} dN/dy|_{y=0}$ for $J/\psi$ per binary collision  ($\langle nb\rangle$) in p-p and Au+Au, as a function of centrality (number of participants). The flat line is the best fit binary-scaling value and the curve is for normal nuclear absorption.\cite{PXe}  }
\label{fig:AuAuJpsi}
\end{figure}

	The $J/\Psi$ data in Au+Au are more a proof of principle than a physics result. The minimum bias data for $4 {\mu b}^{-1}$ integrated luminosity (Fig.~\ref{fig:AuAuJpsi}) show a peak of ~10 $J/\Psi\rightarrow e^{+} + e^{-}$  
events over the mixed-event background. This is rather less than the $\sim 200,000$ events used in the NA50\cite{NA50} ``anomalous suppression'' measurement at CERN. For the next RHIC Au+Au run, we expect a factor of 100 increase of integrated luminosity, 300--400 ${\mu b}^{-1}$, and an additional factor of 10 from the dimuon measurement for a total of $\sim 10000$ $J/\Psi\rightarrow e^+ +e^-$, $\mu^+ + \mu^-$ events. In the interim, we have sliced our 10 events into 3 centrality bins and calculated $B_{ee} dN/dy|_{y=0}$ per binary collision, Fig.~\ref{fig:AuAuJpsi}(right). The data are inconclusive to distinguish between the binary-scaling or standard nuclear absorption curves shown, so we will have to wait for the next Au+Au run at RHIC for more statistics. As the $\Upsilon\rightarrow e^+ +e^-$ is a factor of 1000 down in cross section from the $J/\Psi$, a luminosity upgrade is the only way to get at $\Upsilon$ physics at RHIC. 
 
\section{Conclusions} 
The proof of principle has been established for $J/\Psi$ measurement in Au+Au collisions at RHIC. A factor of 100-1000 more data is needed for a decent measurement, which is expected in the next run. A significant measurement of the $J/\Psi$ $d\sigma/dy$ and total cross section in p-p collisions has been made with only 60 events, with better measurements to follow. Charm has been measured for the first time in RHI collisions via a prompt $e^{\pm}$ signal and 
indicates point-like scaling in contrast to the suppression in high $p_T$ $\pi^0$ production. This is suggestive of a difference in the interaction of light and heavy quark jets with the hot, dense and possibly deconfined medium produced in Au+Au collisions at RHIC. A direct photon measurement awaits improvement of the systematic error in the inclusive photon spectrum. In sum, a very successful start of what we expect should be a long and fruitful program of lepton and photon physics at RHIC.  

\end{document}